\documentclass[prl,aps,twocolumn,showpacs,superscriptaddress]{revtex4-1}

\usepackage{epsfig}
\usepackage{natbib}

\begin{document}

\title{Nearly-critical spin and charge fluctuations in KFe$_2$As$_2$ observed
by high-pressure NMR}

\author{P. S. Wang}
\affiliation{Department of Physics, Renmin University of China, Beijing
100872, China}
\author{P. Zhou}
\affiliation{Department of Physics, Renmin University of China, Beijing
100872, China}
\author{J. Dai}
\affiliation{Department of Physics, Renmin University of China, Beijing
100872, China}
\author{J. Zhang}
\affiliation{School of Energy, Power and Mechanical Engineering, North China
Electric Power University, Beijing 102206, China}
\author{X. X. Ding}
\affiliation{National Laboratory of Solid State Microstructures and Department
of Physics, Innovative Center for Advanced Microstructures, Nanjing University,
Nanjing 210093, China}
\author{H. Lin}
\affiliation{National Laboratory of Solid State Microstructures and Department
of Physics, Innovative Center for Advanced Microstructures, Nanjing University,
Nanjing 210093, China}
\author{H. H. Wen}
\affiliation{National Laboratory of Solid State Microstructures and Department
of Physics, Innovative Center for Advanced Microstructures, Nanjing University,
Nanjing 210093, China}
\author{B. Normand}
\affiliation{Department of Physics, Renmin University of China, Beijing
100872, China}
\author{R. Yu}
\affiliation{Department of Physics, Renmin University of China, Beijing
100872, China}
\affiliation{Department of Physics and Astronomy, Collaborative Innovation
Center of Advanced Microstructures, Shanghai Jiao Tong University, Shanghai
200240, China}
\author{Weiqiang Yu}
\email{wqyu\_phy@ruc.edu.cn}
\affiliation{Department of Physics, Renmin University of China, Beijing
100872, China}
\affiliation{Department of Physics and Astronomy, Collaborative Innovation
Center of Advanced Microstructures, Shanghai Jiao Tong University, Shanghai
200240, China}

\date{\today}


\begin{abstract}
We report a high-pressure $^{75}$As NMR study on the heavily hole-doped iron
pnictide superconductor KFe$_2$As$_2$ ($T_c \approx$ 3.8 K). The low-energy
spin fluctuations are found to decrease with applied pressure up to 2 GPa,
but then increase again, changing in lockstep with the pressure-induced
evolution of $T_c$. Their diverging nature suggests close proximity to a
magnetic quantum critical point at a negative pressure of $P \simeq - 0.6$
GPa. Above 2.4 GPa, the $^{75}$As satellite spectra split below 40 K,
indicating a breaking of As site symmetry and an incipient charge order.
These pressure-controlled phenomena demonstrate the presence of nearly-critical
fluctuations in both spin and charge, providing essential input for the origin
of superconductivity.
\end{abstract}

\maketitle

The presence of multiple active orbitals in iron-based superconductors (FeSCs)
\cite{Hosono_Jacs_130_3296, Chen_PRL_100_247002, Chen_Nature_453_761,
Ren_CPL_12_105} causes a range of highly non-trivial characteristics.
The combination of Fermi-surface nesting, Hund coupling, and Coulomb
interactions produces a rich variety of electronic and magnetic properties
\cite{Dai_NP_8_709} and a complex interplay among the lattice structure,
magnetism, and superconductivity \cite{Paglione_Nature,Stewart_RMP_2011}.

Heavily doped FeSCs, whose Fermi surfaces are quite different from optimally
doped materials, challenge the existing understanding. KFe$_2$As$_2$ has large
hole doping (0.5 hole/Fe), but far from being a regular metal it shows
heavy-fermion characteristics below a low coherence temperature of order 60
K \cite{LiSY_09102806} and superconductivity at a low but finite $T_c$ of 3.8
K \cite{Rotter_PRL_101_107006, Chen_EPL_85_17006}. The absence of electron
pockets around ($\pi$, $\pi$) \cite{Sato_PRL_103_047002} suggests that spin
fluctuations from interband nesting are unlikely, but low-energy electronic
correlations are surprisingly strong. Similarly strong low-energy spin
fluctuations \cite{Zhang_prb_81,Lee_PRL_2011} and significant electron mass
enhancement (average $m^*/m_e \sim 12$) \cite{Kimata_PRL_2011,Storey_PRB_2013}
have been reported in KFe$_2$As$_2$, and appear to strengthen with $x$ in
Ba$_{1-x}$K$_x$Fe$_2$As$_2$ \cite{Storey_PRB_2013}. Controversy over
$s^{\pm}$-wave (favored by interband scattering) or $d$-wave pairing symmetry
\cite{Fukazawa_JPSJ_78_033704, LiSY_09102806, Shin_Science_2014,
Tafti_NP_9_349_2013} raises questions about the effects of electron
correlations on superconductivity.

Recent high-pressure studies of KFe$_2$As$_2$ discovered an anomalous
reversal of $T_c$, which has a minimum at 1.8 GPa \cite{Tafti_NP_9_349_2013}.
Scenarios proposed to explain this include a change of pairing symmetry
\cite{Tafti_NP_9_349_2013,Terashima_PRB_2014} and a $k_z$ modulation of the
superconducting gap \cite{Taufour_2014}. Although spin fluctuations are
essential to FeSC superconductivity, no measurements under pressure have yet
been reported.

In this Letter, we present a high-pressure study of KFe$_2$As$_2$ by
nuclear magnetic resonance (NMR). The $^{75}$As spectra and spin-lattice
relaxation rate ($1/^{75}T_1T$) are measured under pressures up to 2.42 GPa,
revealing three surprising features. First, $1/^{75}T_1T$ is dominated by
strong low-energy spin fluctuations, suggesting incipient antiferromagnetic
order at a quantum critical point near $-0.6$ GPa. Second, the spin
fluctuations show exactly the same reversal behavior as $T_c$, and indeed
identical evolution at all pressures. Third, a line splitting of the $^{75}$As
satellite spectra below 40 K for pressures above 2.4 GPa indicates a
breaking of four-fold symmetry. This effect is caused by charge order, whose
fluctuations we propose are strong around the $d^{5.5}$ electron filling of
KFe$_2$As$_2$. This emergent charge order is accompanied by the enhancement
of spin fluctuations and hence of $T_c$, demonstrating the importance of
nearly-critical charge fluctuations in heavily hole-doped FeSCs.

Our KFe$_2$As$_2$ single crystals were synthesized by the self-flux method
\cite{Terashima_JPSJ_78_063702}. We measure very large residual resistivity
ratios of 1390, indicating extremely high sample quality. We performed
high-pressure NMR measurements using a NiCrAl clamp cell, which reaches a
maximum pressure of 2.42 GPa at $T = 2$ K; to obtain a maximally hydrostatic
pressure, we chose Daphne oil as the medium. The actual pressure was calculated
from the $^{63}$Cu nuclear quadrupole resonance (NQR) frequency of Cu$_2$O
powders \cite{Reyes_Cu2O}. $P(T)$ changes negligibly below 150 K and here we
use the pressure values measured at $T = 2$ K. The superconducting transition
temperature, $T_c$, was determined from the RF inductance of the NMR coil. The
$^{75}$As NMR and NQR signals were measured by the spin-echo technique. All NMR
data were taken with a magnetic field of 10.6 T, which is well beyond $H_{c2}$,
applied in the crystalline $ab$-plane. The spin-lattice relaxation rate
$1/^{75}T_1$ was measured by the spin-inversion method, and the spin recovery
is fitted well by a single $T_1$ component at all pressures, indicating high
sample homogeneity.

\begin{figure}
\includegraphics[width=7.0cm, height=5.6cm]{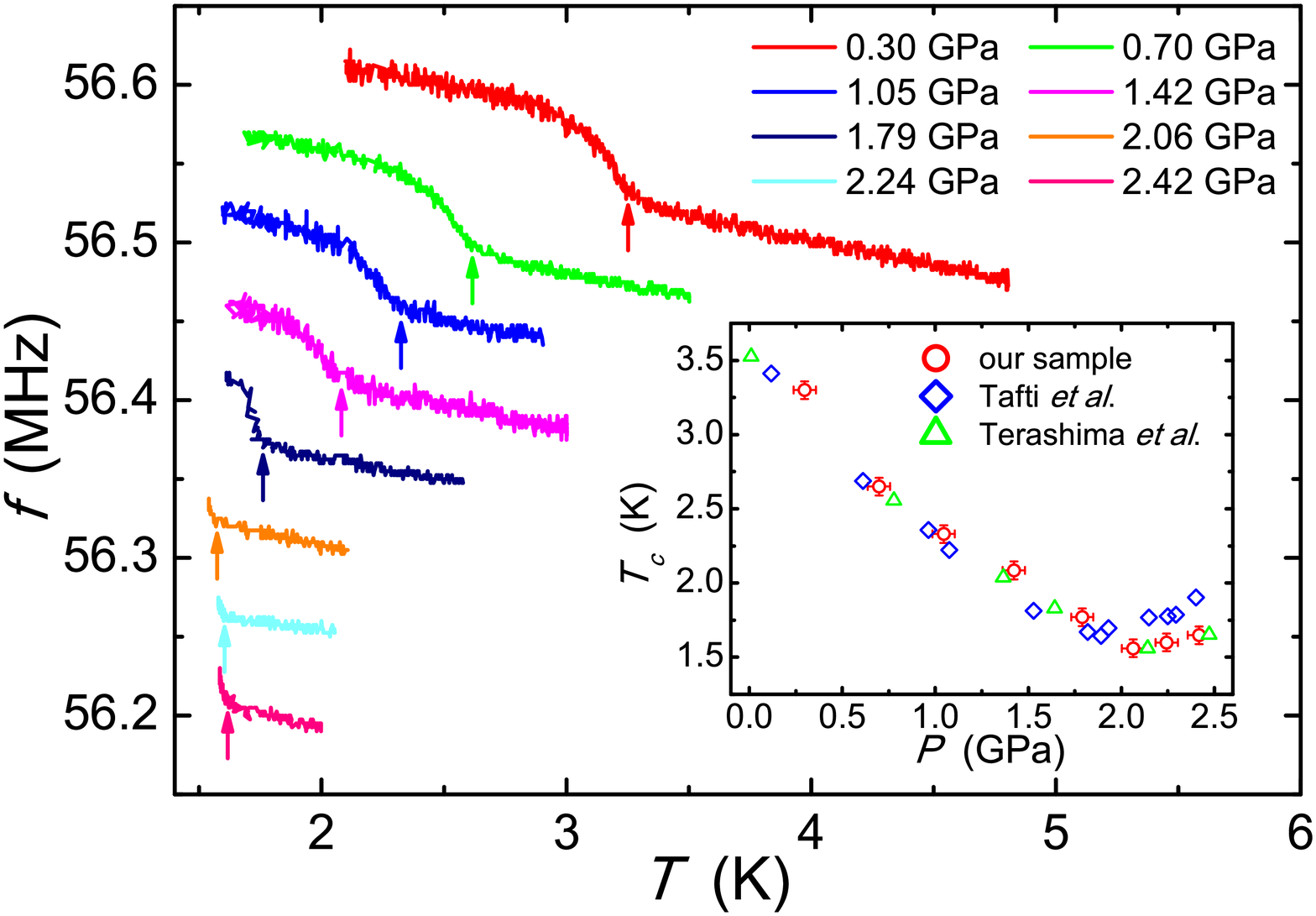}
\caption{\label{tc1}(color online) Resonance frequency $f$ of the detuned NMR
coil at different pressures, as a function of temperature, measured at zero
field. Lines are offset vertically for clarity and arrows indicate $T_c$.
Inset: measured $T_c$ as a function of pressure (open circles), compared with
$T_c$ data reported from transport measurements \cite{Tafti_NP_9_349_2013}
(open diamonds) and from $ac$ susceptibility measurements
\cite{Terashima_PRB_2014} (open triangles). }
\end{figure}

$T_c$ data for KFe$_2$As$_2$ under pressure are determined from the resonance
frequency $f$ of the detuned NMR circuit on cooling. The superconducting
transition causes a sharp increase of $f$ [Fig.~\ref{tc1}] due to the decrease
of coil inductance. $T_c$ as a function of pressure (inset, Fig.~\ref{tc1})
has an initial decrease, at a rate $dT_c/dP \simeq - 1.6$ K/GPa, but reaches a
minimum of $T_c = 1.5 \pm 0.05$ K at $P \simeq 2.1$ GPa, then increases slowly
($dT_c/dP \simeq 0.3$ K/GPa). Our data for $T_c(P)$ are fully consistent
with $ac$ susceptibility measurements \cite{Terashima_PRB_2014}, whereas
transport measurements \cite{Tafti_NP_9_349_2013} found a lower reversal
pressure of $P \approx 1.6$ GPa and a larger $T_c$ beyond this. We suggest
that our results and those of Ref.~\cite{Terashima_PRB_2014} represent more
closely the bulk $T_c(P)$, whereas transport measurements are more susceptible
to boundary superconducting phases in multi-domain samples and to
non-hydrostatic pressure conditions \cite{Yu_PRB_2009}.

\begin{figure}
\includegraphics[width=7cm, height=8cm]{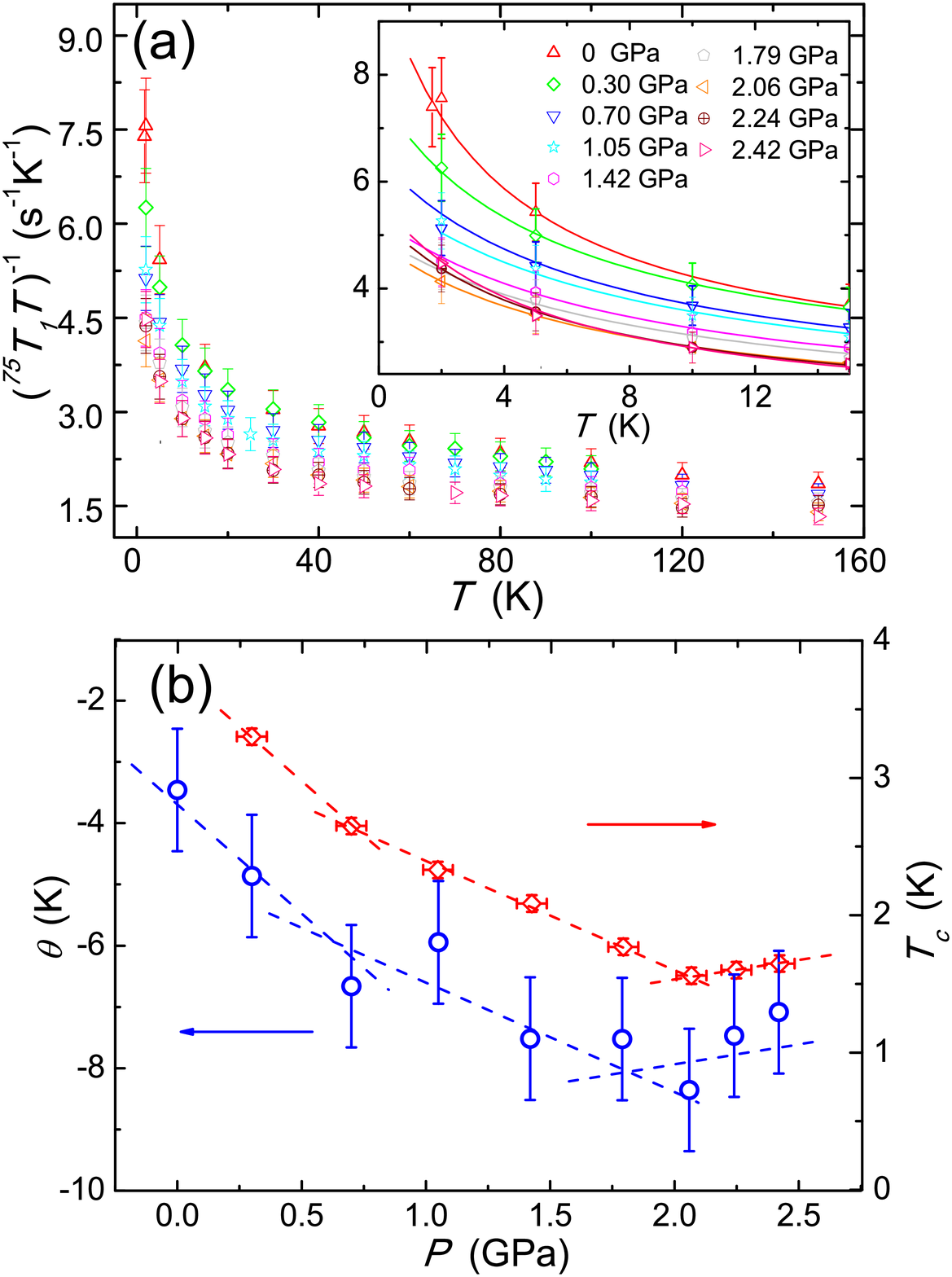}
\caption{\label{invt1t3}(color online) (a) Normal-state $1/^{75}T_1T$ as
a function of temperature from 2 K to 150 K, measured at different
pressures. Inset: zoom of low-temperature data. Solid lines are fits
with the Curie-Weiss form $1/^{75}T_1T = a + b/(T - \theta)$. (b) Values
of $\theta$ extracted from $1/^{75}T_1T$ (left axis) and of the measured
$T_c$ (right axis) as functions of pressure.}
\end{figure}

\begin{figure}
\includegraphics[width=8.5cm, height=6.6cm]{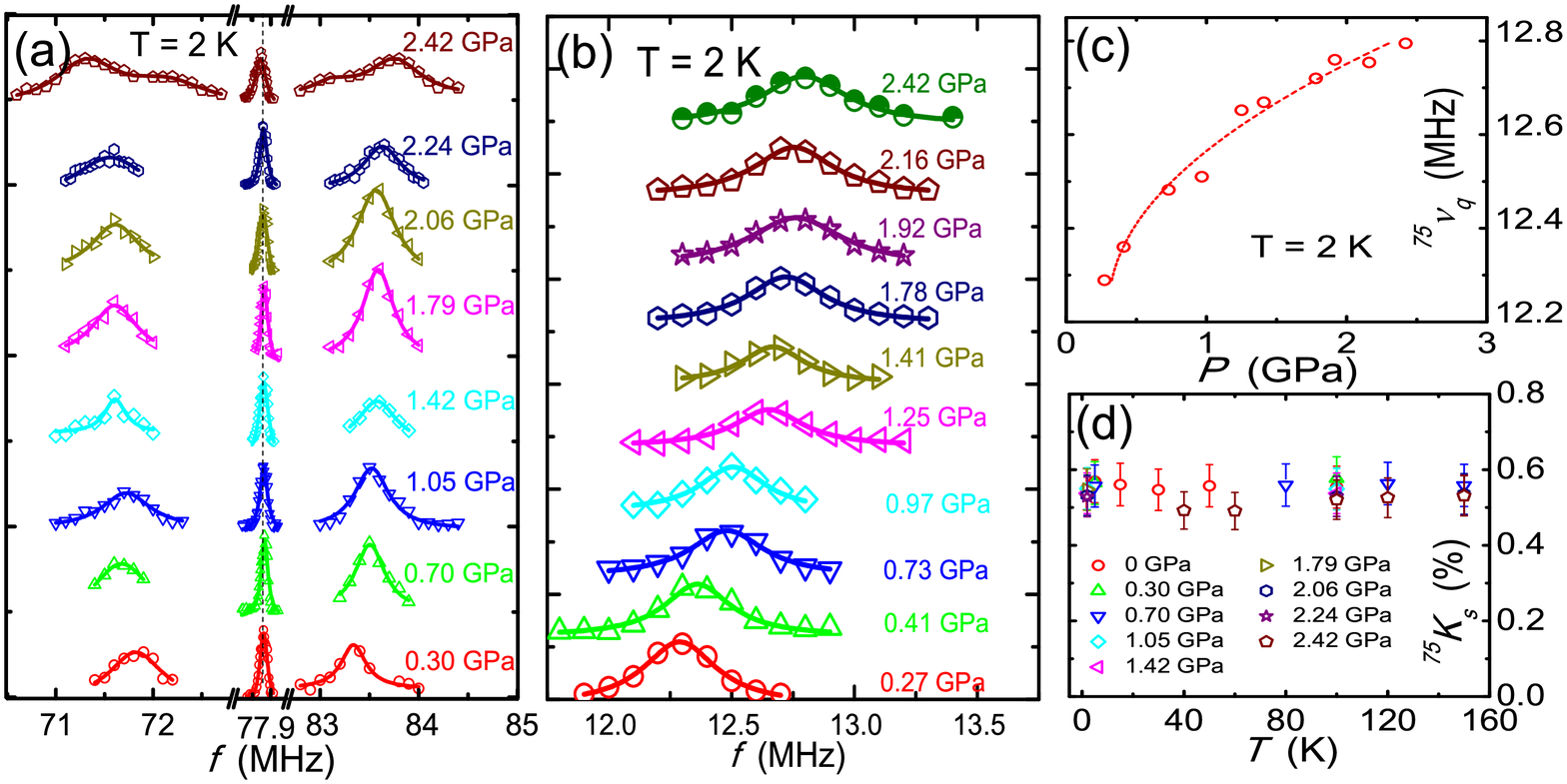}
\caption{\label{spec2}(color online) (a) $^{75}$As spectra measured at
different pressures under a field of 10.6 T applied in the $ab$-plane and
at $T = 2$ K. Data for different pressures are offset for clarity. (b)
$^{75}$As NQR spectra measured at different pressures in zero field and
at $T = 2$ K (data again offset). (c) $^{75}\nu_q$ as a function of pressure.
(d) $^{75}$As Knight shift as a function of temperature. }
\end{figure}

Figure \ref{invt1t3}(a) shows $1/^{75}T_1T$ for a range of pressure values.
Qualitatively, it rises very strongly on cooling, suggesting close proximity
to a magnetic ordering transition. Quantitatively, we obtain an excellent fit
to the low-temperature data ($T \le 40$ K) with the function $1/T_1T = a +
b/(T - \theta)$ [inset, Fig.~\ref{invt1t3}(a)], where $a$ is a
temperature-independent contribution characteristic of itinerant electrons
and the $b$ term is a Curie-Weiss function describing contributions from
spin fluctuations \cite{Moriya}. This behavior contrasts starkly
with the weak temperature-dependence of the Knight shift, $^{75}K$ [shown in
Fig.~\ref{spec2}(d)]. Because $^{75}K$ measures the susceptibility at $q = 0$,
whereas $1/^{75}T_1T$ has contributions from all ${\bf q}$, Fig.~\ref{invt1t3}(a)
indicates that the spin fluctuations are almost exclusively antiferromagnetic
\cite{Imai_prl_102_177005,Kita_JPSJ_77_114709}.

To describe the evolution of $1/^{75}T_1T$ with increasing pressure, we find
that $a$ decreases only slightly, suggesting a small reduction in density of
states on the Fermi surface. Changes of $b$ are very weak, and hence the
low-energy spin fluctuations are characterized almost exclusively by the
Curie-Weiss temperature $\theta$. The extracted value of $\theta$ is small
and negative at all pressures [Fig.~\ref{invt1t3}(b)], with an initial rapid
decrease from $-3.5$ K, at $-5.3$ K/GPa, whose rate slows until a minimum
is reached at 2 GPa; here $\theta$ reverses and increases slowly (1 K/GPa),
i.e.~spin fluctuations are suppressed by low pressures but are enhanced again
at high pressures [Fig.~\ref{invt1t3}(a)].

Figure \ref{invt1t3}(b) contains two key messages. First, from the initial
value of $d\theta /dP \simeq -5.3$ K/GPa, one may estimate that a magnetic
quantum phase transition occurs at negative pressure ($P_c = - 0.6$ GPa).
Such a low $P_c$ implies that magnetic order could be achieved by chemical
pressure. The small value of $\theta$ at ambient pressure reflects that
$1/^{75}T_1T$ is dominated by nearly-critical magnetic fluctuations, which
is a major surprise in a system as heavily hole-doped as KFe$_2$As$_2$.
The initial decrease of $\theta$ under pressure is consistent with cyclotron
resonance measurements, which found a correspondingly reduced mass enhancement
\cite{Terashima_PRB_2014}. CsFe$_2$As$_2$ is a material with a lower chemical
pressure than KFe$_2$As$_2$ (the larger Cs ions expand the in-plane lattice
parameters) and shows a larger mass enhancement \cite{Loehneysen}, in
agreement with our deduction, but no magnetic order. Although $T_c(P)$ is
qualitatively identical to that in KFe$_2$As$_2$ \cite{Tafti_PRB_2014}, the
lower $T_c$ values and lower reversal pressure are not understood. The
second key message is that $\theta$ and $T_c$ change in lockstep as functions
of pressure; such behavior has been observed in other FeSCs \cite{Ji_prl_2013}
and has the unambiguous interpretation that magnetic fluctuations are the
primary contributor to superconducting pairing.

Next we consider the $^{75}$As NMR spectra under pressure. For a nuclear
spin of $I = 3/2$, each spectrum [Fig.~\ref{spec2}(a)] consists of one center
line, whose width is controlled by second-order corrections from electric-field
gradients (EFGs), and two broad satellites subject to first-order corrections.
At $T = 2$ K, increasing pressure causes the satellites to shift further from
the center line, a trend confirmed by the quadrupole frequency, $^{75}\nu_q$,
which is determined by NQR at zero field [Fig.~\ref{spec2}(b)]. $^{75}\nu_q$
increases monotonically with pressure [Fig.~\ref{spec2}(c)], but at a
decreasing rate. Because $^{75}\nu_q$ measures the local EFG at the $^{75}$As
site, this result is consistent with the gradual decrease of the in-plane
lattice parameter under pressure reported by x-ray measurements
\cite{Tafti_PRB_2014}. The Knight shift, deduced from the center
lines, is almost completely constant for all pressures and temperatures
[Fig.~\ref{spec2}(d)], indicating that the Fermi surface and the $q = 0$
spin fluctuations undergo minimal changes at these pressures.

\begin{figure}
\includegraphics[width=7.2cm, height=6.2cm]{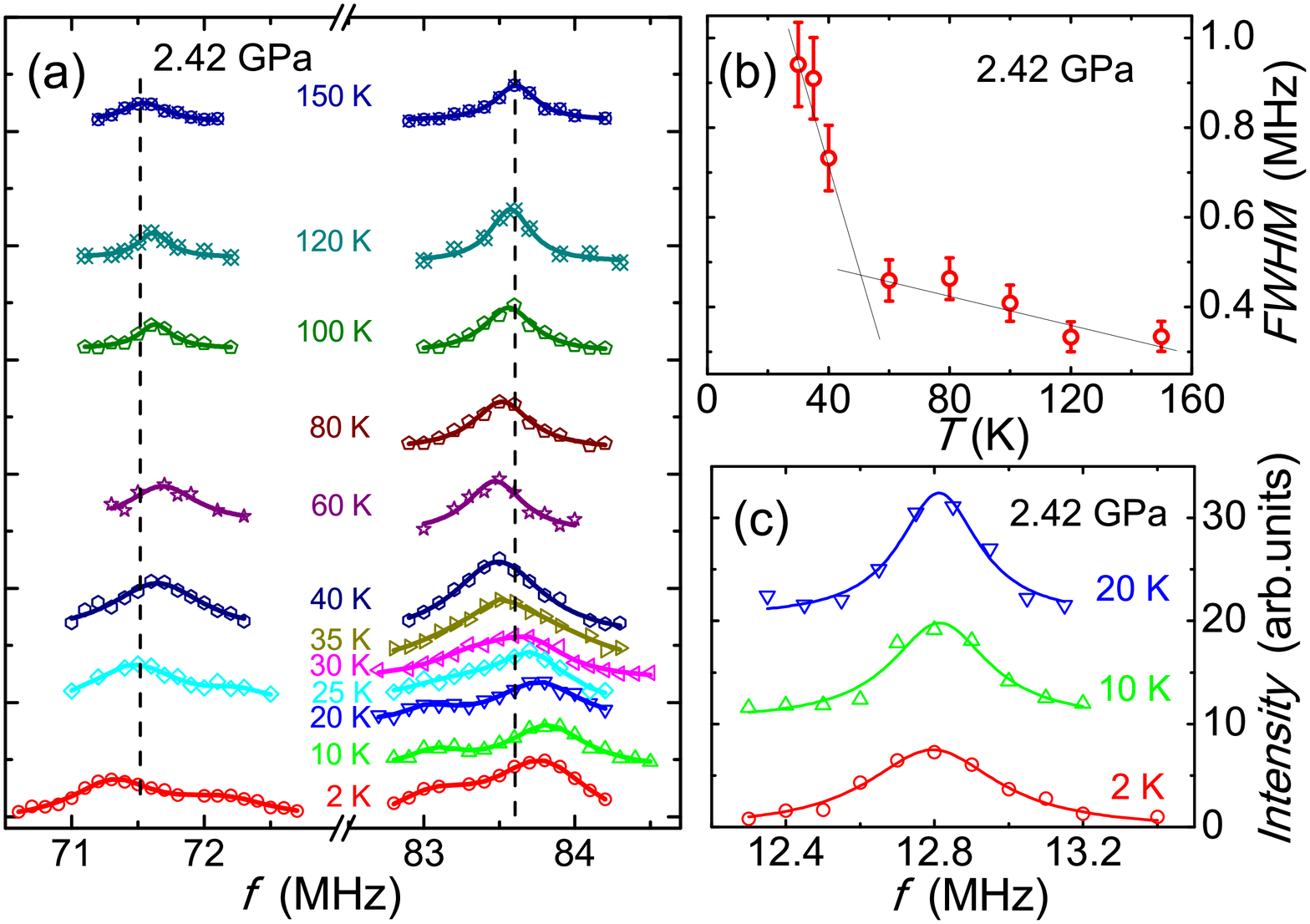}
\caption{\label{struc5}(color online) (a) $^{75}$As NMR satellite spectra
at the highest pressure $P = 2.42$ GPa, measured at different temperatures.
(b) FWHM of the $^{75}$As satellite (centered at 83.5 MHz) as a function of
temperature. (c) $^{75}$As NQR spectra measured at low temperatures with
$P = 2.42$ GPa.}
\end{figure}

Below 2.24 GPa, there is little change in line shapes [Fig.~\ref{spec2}(a)].
At 2.42 GPa, however, each satellite broadens and splits unambiguously into
two peaks with a separation of 1 MHz at $T = 2$ K. By measuring the $^{75}$As
satellite spectra at 2.42 GPa for different temperatures [Fig.~\ref{struc5}(a)],
we observe that each satellite is single-peaked at high $T$ but splits into two
peaks between 30 and 40 K. The full width at half maximum height (FWHM) of the
83.5 MHz satellite at 2.42 GPa, shown in Fig.~\ref{struc5}(b), exhibits a
pronounced kink below $T \approx 40$ K, which marks the onset of splitting.
Because Daphne oil solidifies above 150 K, a sharp line-splitting below 40 K
can only be due to an intrinsic phase transition; simple strain effects or
contributions from uniaxial pressure components could contribute to a
line-broadening, but not to a splitting.

Such a broadening is indeed observed for the center line, whose FWHM increases
from 80 kHz to 100 kHz; any splitting of this line could not exceed 20 kHz.
The absence of measurable splitting in the center line, by comparison with the
1 MHz splitting of the satellites [Fig.~\ref{struc5}(b)], leads us to conclude
that the latter is a charge effect, rather than a magnetic effect. A double
peak in the local EFG indicates two types of local charge distribution at
different As sites, implying a charge ordering on NMR time scales. However, the
$^{75}$As NQR spectra at zero field [Figs.~\ref{spec2}(b) and \ref{struc5}(c)]
remain single-peaked at low temperatures, demonstrating that the amplitude of
the EFG is the same on all As sites. We deduce that the satellite splitting is
caused by a broken in-plane (four-fold) symmetry of the EFG, creating two types
of local charge environment with orthogonal orientations, which respond
differently to the (fixed, in-plane) magnetic field.

Broken in-plane symmetry of the EFG should indicate a structural, orbital,
or charge transition. In underdoped FeSCs, $^{75}$As satellite splitting with
an in-plane field is observed just below the tetragonal to orthorhombic
structural transition, where it is caused by twinning \cite{Ma_NaFeAs}.
However, $x$-ray diffraction measurements find no evidence of a structural
transition in KFe$_2$As$_2$ in this pressure range \cite{Tafti_NP_9_349_2013}.
Thus charge ordering appears the most viable candidate to account for our
observations.

\begin{figure}
\includegraphics[width=7cm, height=3cm]{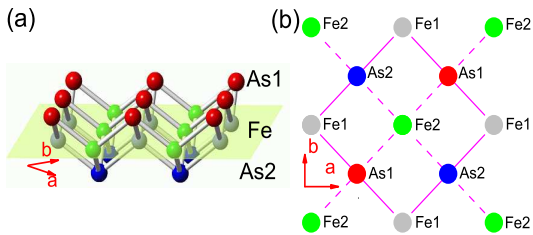}
\caption{\label{co}(color online) (a) Schematic representation of the FeAs
plane with checkerboard charge order on the Fe sites. (b) Planar projection
showing charge-rich Fe1 and charge-poor Fe2 sites breaking four-fold symmetry
at the As sites.}
\end{figure}

For a qualitative understanding of our results, we note that KFe$_2$As$_2$
has a hole doping concentration $x = 0.5$. Such half-doped configurations are
generically less stable, and are often susceptible to charge-disproportionation
fluctuations. These ``mixed-valence'' fluctuations are well known in manganites
to stabilize the CE phase, which has a combined charge, orbital, and spin
order \cite{CE}, and a similar state has been proposed in a two-band FeSC
model \cite{SCO}. In LiV$_2$O$_4$ ($d^{1.5}$), valence fluctuations produce
remarkable heavy-fermion behavior and highly anomalous electronic and magnetic
properties \cite{LVO1,LVO2,LVO3}. We propose a checkerboard charge order on
the Fe sites, shown in Fig.~\ref{co}, where equally charge-rich and charge-poor
sites alternate in two dimensions. The crystal structure of KFe$_2$As$_2$
naturally contains two Fe sublattices with this pattern, because of the two
inequivalent As sites, which in principle supports a checkerboard order. As
a result, the four-fold symmetry of the charge density and thus of the EFG is
broken on the As sites.

We stress that checkerboard charge order is the minimal scenario, as many
different types of charge or orbital ordering with the compatible broken
symmetry are possible. FeSCs are complex, multi-orbital systems where Hund
coupling, on-site repulsion, Fermi-surface nesting, and orbital-selective
correlation effects \cite{Medici_prl_102,YuR_2013} may all compete to induce
order. Next-neighbor Coulomb interactions may be of particular importance for
charge fluctuations in the pattern of Fig.~\ref{co}. More generally, while
checkerboard order provides an appealing real-space illustration of the
mixed-valence phenomenon, in a multiband system it may take a rather different
form with as few as one of the bands actually at commensurate filling.

Although checkerboard charge order breaks a local $C_4$ rotational
symmetry at the As sites (Fig.~\ref{co}), it is quite different from
the concept of nematicity in FeSCs, where anisotropic electronic properties
are usually associated with an orthorhombic structural distortion
\cite{Fisher_Science_2010}. The checkerboard pattern does not drive
orthorhombicity, a result consistent with Ref.~\cite{Tafti_PRB_2014} and
with the lack of anisotropy in any other static observables. Concerning
the nature of the charge-ordering transition, transport measurements
find no additional charge localization in KFe$_2$As$_2$ under pressure
\cite{Taufour_2014}. We suggest that the multiband system (transport
dominated by itinerant bands) and the orbital-selective correlation
effects mask any evidence of charge localization.

Regarding the time scale of charge-ordering, the symmetry-breaking at
$P = 2.42$ GPa appears static on the NMR time scale. However, we cannot
exclude that this is only a quasi-static local order, which is pinned by a
non-hydrostatic component in the applied pressure at its upper limit. As
above, the charge order we observe in the satellite splitting has no other
detectable experimental effects. Thus the question of static order, where
the charge discrepancy $\delta_i$ on each site satisfies $\langle \delta_i
\rangle \ne 0$, or a quasi-static order with $\langle \delta_i \rangle = 0$
but $\langle \delta_i^2 \rangle \ne 0$, remains open. In either case, charge
ordering appears to enhance the spin fluctuations, setting in around 2.4 GPa
while $\theta$ increases above 2 GPa. Although no charge order is detected at
ambient pressure, nearly-critical charge fluctuations ($\langle \delta_i^2
\rangle \ne 0$) may exist over a wide pressure range in KFe$_2$As$_2$ and
contribute strongly to the nearly-critical spin fluctuations we observe.

These spin fluctuations are found \cite{Lee_PRL_2011} to be incommensurate,
exhibiting peaks at $(\pi (1 \pm \epsilon), 0)$ with $\epsilon \simeq 1/6$,
so are quite different from the $(\pi,0)$ antiferromagnetic order of
iron pnictides near $x = 0$ \cite{Dai_Nature_453_899}. Charge order in
the $d^{5.5}$ state should cause significant modification of local magnetic
interactions and the incommensurability may result from stronger frustration
on half of the next-neighbor bonds in Fig.~\ref{co}. If electron filling is
changed from $d^{6}$ to $d^{5}$, the $d_{xy}$ orbitals are thought to account
for the observed increase in electronic correlations \cite{Kimata_PRL_2011,
Storey_PRB_2013}. When a multiband model is used to discuss magnetic order,
$d^{6}$ filling is found at the mean-field level to favor $(\pi,0)$ order
whereas $d^5$ favors a checkerboard [$(\pi,\pi)$] pattern
\cite{Calder_PRB_2012}. KFe$_2$As$_2$ is close to neither filling and strong
(nearly-critical) incommensurate fluctuations are entirely consistent with
strongly competing states of different charge and spin order. We comment
that a scenario of charge-ordering-induced magnetism may emerge from the
correlation-driven scenario \cite{Medici_prl_102,YuR_2013} at $x = 0.5$.

Turning to the connection of charge and spin fluctuations to superconductivity,
Fig.~\ref{invt1t3}(b) shows that $T_c(P)$ and $\theta(P)$ have almost identical
forms, and the pressures for minimum $T_c$ and minimum spin fluctuations
coincide within the error bar. This remarkable positive correlation is hard
to interpret in any way other than spin fluctuations driving the change of
$T_c$ and hence a strong magnetic contribution to superconductivity. Our
results suggest that charge fluctuations may enhance spin fluctuations, as
the origin of the effects above 2 GPa, and thus are also constructive
contributors to superconductivity.

A very similar simultaneous optimization of superconductivity and spin
fluctuations was observed in NaFe$_{0.96}$Co$_{0.06}$As \cite{Ji_prl_2013}.
This electron-overdoped material is close to good Fermi-surface nesting
\cite{Feng_DL_NaFeAs} and the spin fluctuations from itinerant electrons are
thought to play the dominant role in modifying $T_c$. Because KFe$_2$As$_2$
lacks electron pockets and viable nesting options, its spin fluctuations
most likely originate in localized orbitals, which is fully consistent with
weak pairing interactions and the low $T_c$. Valuable insight into the
superconducting state could be obtained by ascertaining whether the
pairing symmetry changes at the pressure-induced reversal in $T_c$
\cite{Tafti_NP_9_349_2013,Terashima_PRB_2014,Taufour_2014}, but we
were unable to study this here.

Finally, we comment that the presence of strong charge and spin fluctuations
is well known in unconventional superconductors, including many underdoped
cuprates \cite{Kivelson_RMP,WuT_nature}, sodium cobaltates \cite{Imai_CO},
and low-dimensional organic systems \cite{OS}. Thus it is not a complete
surprise that a $d^{5.5}$ material such as KFe$_2$As$_2$ shows similar
strong-correlation effects.

In summary, we have performed high-pressure $^{75}$As NMR measurements on
the heavily hole-doped iron superconductor KFe$_2$As$_2$. We find strong
low-energy spin fluctuations indicating close proximity to a hidden magnetic
ordering at low negative pressures. We find a charge order at high
pressures, suggesting the predominance of nearly-critical charge (valence)
fluctuations over much of the phase diagram. This is consistent with the
low-$T$ heavy-fermion behavior, non-Fermi-liquid behavior at intermediate
$T$, and low $T_c$ value. We find that the spin fluctuations have a
non-monotonic pressure dependence identical to that of $T_c$, reinforcing
that pairing is of magnetic origin. We suggest that the half-integer
filling ($d^{5.5}$) causes nearly-critical charge fluctuations that are
intrinsically linked to the nearly-critical spin state and hence to
unconventional superconductivity.

Work at Renmin University of China is supported by the National 
Science Foundation of China (NSFC) under Grant Nos.~11174365, 11222433,
11374361, and 11374364, the National Basic Research Program of China (NBRPC)
under Grant Nos.~2011CBA00112 and 2012CB921704, the Fundamental Research Funds
for the Central Universities, and the Research Funds of Renmin University of
China. 
Work at Nanjing University is supported by the NSFC
and by the NBRPC (Grant Nos. 2011CBA00102 and 2012CB821403).

\end{document}